\documentclass[acmsmall,screen]{acmart}
\AtBeginDocument{%
  }

\setcopyright{none}
\copyrightyear{2026}
\acmYear{2026}

\acmJournal{JACM}

\usepackage{algorithmic}
\usepackage{graphicx}
\usepackage{textcomp}
\usepackage{listings}
\usepackage{multirow}
\usepackage{color}
\usepackage{tcolorbox}
\tcbuselibrary{skins}
\usepackage{url}
\usepackage{subcaption}

\definecolor{codegreen}{rgb}{0,0.6,0}
\definecolor{codegray}{rgb}{0.5,0.5,0.5}
\definecolor{codepurple}{rgb}{0.58,0,0.82}
\definecolor{backcolour}{rgb}{0.95,0.95,0.92}

\lstdefinestyle{codestyle}{
  backgroundcolor=\color{backcolour}, commentstyle=\color{codegreen},
  keywordstyle=\color{magenta},
  numberstyle=\tiny\color{codegray},
  stringstyle=\color{codepurple},
  basicstyle=\ttfamily\footnotesize,
  breakatwhitespace=false,         
  breaklines=true,                 
  captionpos=b,                    
  keepspaces=true,                 
  numbers=left,                    
  numbersep=5pt,                  
  showspaces=false,                
  showstringspaces=false,
  showtabs=false,                  
  tabsize=2
}

\newcounter{finding}

\newenvironment{finding}{
  \refstepcounter{finding}
  \begin{tcolorbox}[
    colback=gray!10,
    colframe=gray!50,
    boxrule=0.4pt,
    arc=1pt,
    left=6pt,
    right=6pt,
    top=4pt,
    bottom=4pt,
    fontupper=\small,
    before skip=8pt,
    after skip=8pt,
    enhanced
  ]
  \textbf{Finding~\thefinding:}
}{
  \end{tcolorbox}
}

\begin{document}

\title{Can we find bugs using LLM-generated oracles?}

\author{Michael Konstantinou}
\email{michael.konstantinou@uni.lu}
\affiliation{%
  \institution{University of Luxembourg}
  \city{Luxembourg}
  \country{Luxembourg}
}

\author{Renzo Degiovanni}
\affiliation{%
  \institution{Luxembourg Institute of Science and Technology}
  \city{Luxembourg}
  \country{Luxembourg}}
\email{renzo.degiovanni@list.lu}

\author{Mike Papadakis}
\email{michail.papadakis@uni.lu}
\affiliation{%
  \institution{University of Luxembourg}
  \city{Luxembourg}
  \country{Luxembourg}
}

\renewcommand{\shortauthors}{Konstantinou et al.}

\begin{abstract}
Unit testing is vital in software development. Typically, a unit test consists of a test prefix and a \emph{test oracle} which captures the developer's intended behaviour. Traditional test generation tools (e.g. Randoop and Evosuite) often produce oracles that mirror the program’s actual behavior rather than the expected one, limiting their ability to automatically detect bugs as users must manually verify if the generated assertions are correct. Recent approaches leverage Large Language Models (LLMs), trained on vast datasets, to generate developer-like code and test cases. Although successful in generating tests, the question of whether such LLM-generated oracles can automatically find bugs, i.e., expected software behavior, remains unanswered. We conduct a controlled experiment to answer this question, by studying LLMs on two tasks, namely, test oracle classification and generation, and assessing whether LLM oracles capture the actual or the expected behavior. The study includes test cases and oracles written by developers and automatically generated for 24 Java repositories. Our findings show that LLM-based test generation approaches mainly capture the actual program behavior making bug detection difficult. We also find that LLMs are better at generating oracles than classifying them. Notably, LLM-generated oracles have a higher fault detection potential than the Evosuite ones.
\end{abstract}

\maketitle

\section{Introduction}

Software testing is an essential part of the software development, maintenance and evolution cycle as it aims at improving code quality~\cite{AmmannOffutt2016}. Designing test cases is also a time consuming and labour-intense activity that is typically performed manually~\cite{BarrHMSY15}.

To deal with this issue, many automatic test generation approaches utilising various types of technologies, such as evolutionary algorithms~\cite{6004309}, symbolic execution\cite{TillmannDeHalleux2008}, feedback-directed random search~\cite{pacheco2007randoop} and specification mining~\cite{DallmeierKMHZ2010}, have been proposed. These techniques aim at generating test cases that exercise the software under test guided by one or more coverage criteria, i.e., generating test suites that maximize coverage such as branch coverage or mutation testing, with many studies reporting that they achieve comparable or even higher coverage scores than the developer test suites \cite{RojasFA15, Yuan0DW00L24}.    

While effective at covering code (or killing mutants), automatic test generation falls short in finding faults, particularly business-logic related faults. This is because of the inherent inability of these techniques to compose test oracles (test assertions) that capture the expected program behaviour. This means that the fault detection ability of these techniques is limited to zero-level oracles, such as program crashes or memory violations (when applied at system level).

To reveal business-logic software faults with test generation techniques one needs to manually validate and correct, when needed, the generated tests and their respective oracles~\cite{RojasFA15}. In other words, to reveal logic faults, users need to define the expected program behaviour that is tested by the generated tests, which is to be contrasted with the actual program behaviour. This is why test generation approaches, such as Evosuite~\cite{FraserA11} or KLEE~\cite{CadarDE08}, often generate test oracles that capture the actual program behaviour assuming that it is correct, i.e., making the assumption that the current implementation under analysis is correct, which makes them incapable of finding logic-related faults. 

With the advent of language models, many test generation techniques have been proposed that integrate ML techniques as well as powerful large language models (LLMs)~\cite{yuan2023no,schafer2023empirical,WangHCLWW2024}. LLMs have been trained on a huge amount of code and data, and as a result they can generate developer-like code and test cases. The advantage of LLMs, compared to traditional techniques, is that they make use of the natural channel of code~\cite{AllamanisBDS18}, such as naming conventions, and exploit similarities with existing code on which they have been trained to effectively generate tests.

Some approaches employ LLMs to generate only the test oracle, instead of generating the entire test suite. Given a test prefix, these techniques generate a test assertion (test oracle) that hopefully captures the expected program behaviour. The idea is that existing test generation techniques can be combined with the oracle generation ones, to generate tests that reveal software faults. For instance, TOGA \cite{10.1145/3510003.3510141} and TOGLL \cite{hossain2024togllcorrectstrongtest} use Evosuite to generate test prefixes, and leverage LLMs to generate the test oracles. These approaches assume that LLMs can guess the expected program behaviour and thus, use the LLMs to generate correct test oracles. 

In view of this, we investigate the extent to which LLMs are capable of guessing the expected program behaviour in contrast to the actual one, i.e., the extent to which LLMs generate test oracles capturing the expected over the actual program behaviour. An answer to this question in favour of the former case signifies that LLMs can be used to reveal logic-related faults. An answer in favour of the latter case signifies that LLMs are more suited for regression testing. 

We employ four LLMs from different families, namely GPT-3.5-turbo, GPT-4o-mini, DeepSeek V3, and Codestral, to study the behaviour of LLMs on two test oracle tasks proposed in prior work: (i) test oracle classification, which involves judging the correctness of externally provided assertions, and (ii) test oracle generation. The evaluation was conducted using a diverse set of inputs, including developer-written and automatically generated test cases and oracles, as well as both buggy and correct implementations from 24 open-source Java repositories.

Interestingly, our results show that LLMs are more likely to generate test oracles that capture the actual program behaviour (what is actually implemented) rather than the expected one, i.e., the intended behaviour. Additionally, we find that the overall performance of the LLMs is relatively low (less than 50\% accuracy in some cases) meaning that LLMs do not provide a strong oracle correctness signal. Therefore, we conclude that all LLM oracle suggestions need human inspection. 

We also find that LLMs are better at generating test oracles than judging the correctness of externally generated oracles (up to 13.29\% more accuracy on average). LLM-generated test oracles are heavily impacted by the naming conventions used by the tests, they have up to 15.40\% higher performance when using developer-written naming conventions than when using the Evosuite's ones. Interestingly we find that LLMs generated assertions that led up to 2.96\% higher mutation score than Evosuite's test oracles, indicating that they can be used for test augmentation. 

Taken together, our results corroborate the conclusion that unless the code has meaningful test or variable names, LLMs can mainly be used to capture the actual program behaviour (thus to be used for regression testing). Additionally, we find that LLMs could be a good addition to existing test generation tools, or to the test writing task, by using them to perform test augmentation. 

Overall, this work raises awareness of the practical issues involved, advantages and disadvantages of the LLM-based test oracle generation abilities.

Simply put, the paper's contributions are the following:
\begin{enumerate}
    \item An empirical evaluation of the capabilities of LLMs in an assertion classification problem studied under different circumstances and with the usage of different prompt variations.
    \item An empirical evaluation of the capabilities of LLMs in an assertion generation problem studied under different circumstances and with the usage of different prompt variations.
    \item We provide a replication package that includes the two large datasets, based on developer-written and Evosuite-generated test suites and oracles, used in the empirical study at: \url{https://doi.org/10.5281/zenodo.13867480}.
\end{enumerate}

 The paper proceeds by covering the relevant background, followed by a detailed description of the experimental setup. Next, we present the results, accompanied by a discussion based on our findings. We then address the threats to validity of our study and review the related work. Finally, we present the conclusions of our study.

\section{Background}

\subsection{Traditional Test Oracle Generation}

Evosuite \cite{6004309} is a state-of-the-art tool for automated test generation, based on a search-based method. It generates test cases by analyzing the program's implementation and current execution. It suffers from two main limitations though. Its first limitation is that Evosuite's test cases are hard to read and interpret. The second limitation is that Evosuite generates its test cases assuming that the current implementation is correct. Therefore, if the code is buggy, it is very likely that some of the generated tests will be incorrect.

Randoop \cite{pacheco2007randoop} is a feedback-directed random test generation tool. Just like Evosuite, it also produces regression oracles but it also aims to generate tests that violate the relevant API contract(s). Similarly, the limitation of generating tests out of a buggy code is present here as well. 

\hfill

\subsection{Neural networks and LLMs for test oracle generation}

Neural networks are meant to detect patterns in the documentation string (docstring) or the code in order to generate an answer. With proper training, someone can generate code using LLMs without manually writing the rules or the patterns to seek for like in other approaches such as specification mining or rule-based systems. Therefore, a number of approaches have been tried using LLMs for test oracle generation. 

ATLAS \cite{atlas} is a recurrent neural network based approach, which uses a test prefix and a unit under test in order to generate an assertion. It does not use any other information (e.g. docstring) and it focuses only on the inference of assertion test oracles. However, when transformed based models have been used for this purpose, their results outperformed ATLAS \cite{white2020reassertdeeplearningassert} \cite{Tufano_2022} \cite{mastropaolo2021studyingusagetexttotexttransfer}. Those tools however, only focus on the production of assertion oracles and do not attempt to infer or evaluate possible exception oracles. 

TOGA \cite{10.1145/3510003.3510141} is an approach that uses two neural network models based on CodeBERT \cite{feng2020codebertpretrainedmodelprogramming}, and an internal rule-based system in order to generate both: exception and assertion test oracles. The internal mechanism of TOGA initially uses an exception classifier, which will classify whether the given test prefix will throw an exception or not. If the classifier outputs that the test will throw an exception, then manually the tool will generate an exception oracle as typically done using Junit4. Otherwise, TOGA will scan the test and extract a few possible assertions using its own rule-based system. 

Afterwards, a second neural network model is used which will evaluate the correctness of the assertions in order to pick the assertion that is more likely to be correct. Despite its novel approach and its improvement on the exception oracle finding, its internal rule-based system relies a lot on the Evosuite test style convention. For instance, it generates a candidate assertion assuming that the test prefix contains a final line similar to the one used by Evosuite, the one that executes the method under test. Moreover, other studies that evaluated TOGA showed a high number of false positives \cite{10.1145/3611643.3616265} \cite{10.1145/3597926.3598080}. 

Following TOGA, TOGLL \cite{hossain2024togllcorrectstrongtest} is a study that fine-tuned and assessed 7 LLMs for test oracle generation and conducted its evaluation on six different prompt variations. After identifying the best prompt variations and the best performing LLM, they introduced TOGLL, a tool based on their best fine-tuned LLM. It can generate about 3.8 times more correct assertions and 4.9 times more exception oracles than TOGA. Although significantly improved results compared to TOGA, it also suffers from false positives (about 25\% for assertion oracles) and compilation errors since about 5\% of the assertions produced do not compile. In addition, it has a maximum token limit of 600 tokens; which in about 3\% of their cases could not be used on the sixth prompt variation. 

\subsection{Mutation Testing}
\label{sec:mutation-testing}
Mutation testing~\cite{PapadakisK00TH19} is a test adequacy criterion where test requirements are represented by \emph{mutants}, i.e., artificially seeded faults obtained from slight syntactic modifications to an original program (e.g., \texttt{size > 0} is mutated to \texttt{size > 1}). Mutants are used to assess the effectiveness and thoroughness of a test suite, by measuring how many of these artificial faults the suite is able to detect. 
Whether there exists a test case that is capable of producing observable outputs that distinguish between the mutant and the original program, we say the mutant is \emph{killed}; otherwise, the mutant \emph{survived}. 
\emph{Equivalent} mutants are those that cannot be killed as they behave as the original program. 
The \emph{mutation score} (MS) is computed as the ratio between killed mutants over the total number of generated mutants, and gives an estimation of the fault detection capabilities of the test suite.  

In this study we use two mutant generation tools, namely, $\mu$BERT~\cite{degiovanni2022mubert} and PiTest~\cite{ColesLHPV16}, each one for a different research question. 
$\mu$BERT~\cite{degiovanni2022mubert} is a mutation testing tool that uses a pre-trained language model (CodeBERT) \cite{DBLP:conf/emnlp/FengGTDFGS0LJZ20} to generate mutants by masking and replacing tokens. 
Since it works at source code level, we use $\mu$BERT to generate buggy code versions that is used in our experiments to answer RQ1. 
PiTest~\cite{ColesLHPV16} is one of the state-of-the-art mutation testing tools that seeds faults using syntactic transformation rules (aka mutant operators) at the bytecode level. 
We particularly use PiTest in RQ4, to compute and compare the mutation scores between the LLM-generated test oracles and EvoSuite's generated ones.

\section{Experimental Setup}

\subsection{Approach}

The aim of this study is to evaluate the behaviour of LLMs under the possible circumstances that could occur in a real life scenario. Intuitively, on a test generation problem, an LLM could be used to generate tests in a buggy version as well as the correct implementation. Additionally, on an oracle generation problem, the LLM has been used to classify oracles as well as generating oracles. Furthermore, several studies showed how optimizing the prompt can affect the performance of the LLM for such a task. 

Having studied the aforementioned, we conducted several experiments, designed to capture any combination of the following four categories:\\

\begin{enumerate}
    \item Code status (Buggy or Correct)

    The actual implementation may or may not contain bugs. Several traditional approaches such as the use of evolutionary algorithms or symbolic execution, do not typically work under buggy implementations. This study examines the capabilities of LLMs for both buggy and correct code. 
    
    \item Oracle classification and generation

    LLMs have been previously used for oracle classification and oracle generation. The study examines the behaviour of LLMs on both tasks.
    
    \item Prompt variation

    Previous studies showed that the information added to the prompt fed to the LLMs influences the performance of the model. For instance, prompts that contain more information regarding the code under test, tend to yield different results. Therefore, in this study we make use of three different prompt combinations that contain different code information.
    
    \item Code quality

    The code under test may or may not contain descriptive comments and may or may not be automatically generated. The study assesses the performance of LLMs on both scenarios: developer-written test prefixes and automatically generated ones. 
\end{enumerate}

The combination of these four categories, yields a number of cases that we examine individually. Typically, a tool evaluation is described by its total accuracy in all such cases. Since the purpose of this study is to provide further information on the behaviour of LLMs on a test generation task, we review each possible case individually. For instance, the combination of the code status (buggy or non-buggy version) with the three prompt variations results into six in-depth case studies. Hence, instead of providing a unified accuracy measurement, we analyze the accuracy of the LLM on each of the possible cases. 

Combining the above, we designed a controlled experiment that is split across four possible research questions, to provide further insights on how the LLMs behave on a test oracle problem, and whether LLMs can help us find bugs in the first place. 

\subsection{Research Questions}

Our study is designed to assess the performance of LLMs in two different, but related, tasks: test oracle classification~\cite{10.1145/3510003.3510141}, and test oracle generation~\cite{hossain2024togllcorrectstrongtest}. The study also aims to evaluate the fault detection potential of the generated test oracles and identify the conditions under which LLM performance improves. 

We start by checking whether LLMs can capture the program's intended program behaviour or not. Thus, we ask:

\textbf{RQ1: (How accurately) Can LLMs capture the actual or the expected program behaviour when classifying externally defined test oracles?}

To answer to this question, we design a controlled experiment in which we assess if LLMs can correctly classify a given test oracle assertion for a given code under test (ie. a test oracle classification task) under four possible scenarios: 

\begin{itemize}
    \item Correct code and test with correct assertion \textbf{(CC+CA)}
    \item Correct code and test with wrong assertion \textbf{(CC+WA)}
    \item Wrong code and test with correct assertion \textbf{(WC+CA)}
    \item Wrong code and test with wrong assertion \textbf{(WC+WA)}
\end{itemize}

\begin{figure}
    \centering
    \includegraphics[width=1\linewidth]{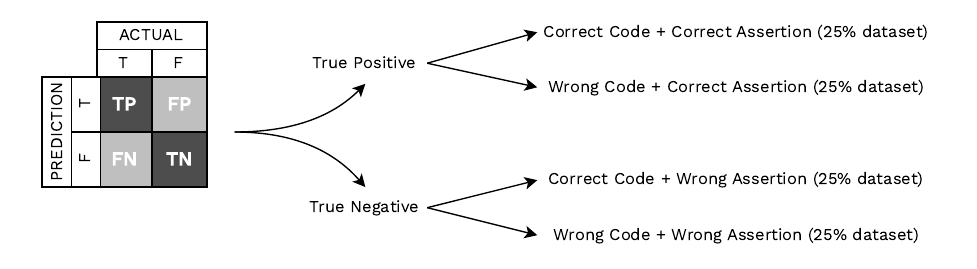}
    \caption{Experimental design: Traditionally, accuracy in classification metrics is measured using combined metrics of true positives, true negatives, false positives and false negatives. However, this is not informative in this case. Therefore, in this study, we break the problem into four equal parts that capture all possible combinations, in order to examine the behaviour of LLMs under different versions.}
    \label{fig:exp_design}
\end{figure}

Each scenario is composed of two parts: the source code of the program under test, and a test case with an oracle assertion. We measure if LLMs can identify the intended behaviour of the code and classify correct assertions as positives, when confronted with clean and buggy code. This is important to avoid the so-called clean program assumption \cite{ChekamPTH17} that is known to impact the performance of test criteria and simulate a realistic scenario, i.e., the case that a buggy code is tested. Assertions not capturing the intended program behaviour should be classified as negatives, even under the presence of correct code. These are also important since they provide wrong developer signal, which in practice would mean that developers will need to investigate false alerts. Figure~\ref{fig:exp_design} illustrates how the evaluation dataset is divided accordingly.

In this study, we distinguish between correct and wrong code, as well as correct and wrong assertions, as defined below. These definitions describe the instances considered throughout our experiments, while Section~\ref{sec:datasets} provides further details on how these instances are extracted.

\begin{itemize}
\item \textbf{Correct code}: The original implementation under test, which is assumed to be free of faults.

\item \textbf{Wrong code}: A modified version of the original implementation containing injected faults. In other words, a buggy version of the current code under test. These faults are introduced via mutation applied to the method under test; therefore, the resulting code is known to produce failing behavior and be relevant with the associated test case.

\item \textbf{Correct assertion}: The original test oracle associated with a test case, which correctly captures the expected behavior. The assertion is syntactically valid and semantically consistent with the expected behaviour of the original implementation. Therefore, the test passes when executed against the original (correct) implementation.

\item \textbf{Wrong assertion}: An alternative oracle associated with the same test case that does not represent the expected behaviour. In this study, a wrong assertion is syntactically valid but semantically incorrect with respect to the expected behaviour of the original implementation. Thus, when the test prefix and the original implementation remain unchanged, replacing the correct assertion with the wrong assertion causes the test to fail.

\end{itemize}

Following previous studies~\cite{yang2024empirical, ouedraogo2024large}, we explore three different prompt variations that include/exclude class method implementations and documentation. We use the dataset from the evaluation of TOGA~\cite{10.1145/3611643.3616265} and TOGLL~\cite{hossain2024togllcorrectstrongtest} to gather Java subjects, and their corresponding test prefixes with correct test oracle assertions, generated by Evosuite. To cover the four aforementioned scenarios, we augment this dataset by running a mutation testing tool, $\mu$BERT~\cite{degiovanni2022mubert}, to generate invalid code versions, and GPT~\cite{brown2020languagemodelsfewshotlearners} to generate invalid test oracle assertions (i.e., those failing when the test is run). 

Intuitively, a high accuracy would reflect a good LLM capability for identifying the intended program behaviour, while on the contrary, a low accuracy would suggest that the LLM is affected by the actual program behaviour. 

While RQ1 studies LLM's sensitivity on a more semantic perspective of the code and test oracles (i.e. fed with correct and buggy versions), now we aim to focus on a more syntactic perspective. Recent studies~\cite{SallouDP2024} have shown that LLM's performance is affected when executed on out of distribution inputs.  Hence, we wonder:\\

\textbf{RQ2: What can influence the finding of an expected oracle instead of an actual oracle?}

We study whether LLM performance is influenced by meaningful test names and descriptive variable names. Previous studies have proposed combining traditional automated testing tools, such as EvoSuite, to generate test prefixes, with LLMs to generate test oracles for detecting bugs~\cite{10.1145/3510003.3510141}. However, such automatically generated tests typically contain generic test and variable names that provide little semantic information. We hypothesize that the absence of this information may limit the ability of LLMs to infer the expected behaviour of the code and, consequently, reduce their effectiveness in generating test oracles. It should be noted that, although previous work has explored using LLMs to produce more meaningful names for EvoSuite-generated tests, such techniques were not employed by prior studies on LLM-based test oracle generation~\cite{deljouyi2025leveraging, daka2017generating}.

Let us consider Figure~\ref{lst:rq2}. The developer-written test contains relevant information that reflects the expected behaviour of the system. For instance, the test specifies that an exception should be thrown when an existing path is provided. In combination with descriptive variable names such as \textit{path}, the LLM is exposed to additional cues about the intended behaviour of the code.

In this research question, we follow the same experimental procedure as in RQ1, but evaluate the performance of the LLMs using developer-written tests. Subsequently, we progressively modify these tests to resemble the naming conventions of EvoSuite-generated test cases, in order to investigate how such transformations affect LLM performance. The experiment includes the following two modifications:

\begin{enumerate}
    \item \textbf{Anonymous names:} Test method names are renamed to the pattern \textit{testX} (e.g., in the previous example, \textit{shouldThrowForExistingPath} is renamed to \textit{test0}).
    
    \item \textbf{Anonymous names + Anonymous variables:} In addition to the previous modification, all variable names are replaced with the pattern \textit{varX} (e.g., in the previous example, \textit{path} is renamed to \textit{var0}, and the variable \textit{conf} is renamed to \textit{var1}, etc.).
\end{enumerate}

\begin{figure}[bt]
\centering

\begin{minipage}{0.48\linewidth}
\begin{lstlisting}[language=Java, caption=Developer written test case]
@Test
void shouldThrowForExistingPath() {
    // given
    final String path = "config.me"; 
    CommentsConfiguration conf = new CommentsConfiguration();
    conf.setComment(path, "Old", "Comments", "1", "2", "3");
    
    // when
    IllegalStateException ex = assertThrows(IllegalStateException.class, () -> conf.setComment(path, "New Comment"));
    
    // then
    assertThat(ex.getMessage(), equalTo("Comment lines already exists for the path 'config.me'"));
    assertThat(conf.getAllComments().keySet(), contains(path));
}
\end{lstlisting}
\end{minipage}
\hfill
\begin{minipage}{0.48\linewidth}
\begin{lstlisting}[language=Java, 
caption=Modified test case (all modifications),
emph={test0,var0,var1,var2},
emphstyle=\bfseries\underline]
@Test
void test0() {
    // given
    final String var0 = "config.me"; 
    CommentsConfiguration var1 = new CommentsConfiguration();
    var1.setComment(var0, "Old", "Comments", "1", "2", "3");
    
    // when
    IllegalStateException var2 = assertThrows(IllegalStateException.class, () -> var1.setComment(var0, "New Comment"));
    
    // then
    assertThat(var2.getMessage(), equalTo("Comment lines already exists for the path 'config.me'"));
    assertThat(var1.getAllComments().keySet(), contains(var0));
}
\end{lstlisting}
\end{minipage}

\caption{Original developer-written case (left) alongside its modified version (right)}
\label{lst:rq2}
\end{figure}

The findings of this experiment will give a new perspective on the usage of LLMs in the generation of test oracles that capture the expected behaviour. Having studied the LLM's performance on classifying test oracles, we investigate their generative performance and ask: 

\textbf{RQ3: (How accurately) Can LLMs generate test oracles capturing the expected program behaviour?}

Using the three different prompt variations as before, we ask the LLM to generate one test assertion. We then evaluate their correctness by incorporating the generated assertions into the given test prefixes and executing them on the program under test. An assertion is considered \emph{invalid} if it fails to compile or if its execution results in a failure or error. Conversely, if the assertion executes successfully, it is classified as \emph{correct}. These findings can provide further information regarding the capabilities of LLMs and how they can achieve their highest potential for test oracle generation. 

However, prior work does not restrict generation to a single test oracle. Instead, LLMs are typically prompted to produce multiple assertions (e.g., three or five), which has been shown to improve performance~\cite{10.1145/3510003.3510141,10804561}. To align with this setting, we repeat the experiment by prompting the LLM to generate five assertions per test case. Following prior work, we denote the accuracy of single-assertion generation as Acc@1 and the accuracy of five-assertion generation as Acc@5.

Finally, we aim to compare the strength of the test oracle assertions generated by the LLM and a traditional test generation technique, Evosuite. Then, we ask:

\textbf{RQ4: How strong are the oracles generated by the LLM?}

The successful execution of a test oracle does not guarantee that it meaningfully constrains or validates the behaviour of the code under test. In particular, overly generic or weak oracles may still allow tests to pass while capturing only trivial or even irrelevant properties. Such oracles may fail to detect faults in the future despite their execution status. To assess the strength of the generated oracles, we evaluate their ability to reveal defects using mutation testing.

To answer to this question, we compute the mutation score of the test suites that integrate Evosuite's and the LLM generated test oracles (Acc@5 setup), respectively. 
We use PiTest~\cite{ColesLHPV16} for this task, since it is very efficient and the mutants are generated at byte code level, and are independent to the ones used in the previous RQs, mitigating any possible bias. 
The answer to this question will provide further insights regarding any potential advantage(s) of using LLMs in test oracle generation problem against traditional methods.

\subsection{Prompt Design}

Several studies evaluating LLMs on test generation concluded that the prompt design affects the performance of the LLM \cite{yang2024empirical, ouedraogo2024large}. When 
evaluating TOGLL, their study included 6 prompt variations. Each prompt included more information regarding the method under test, the test prefix and the documentation string. It has been shown that the prompts that included the method under test and the test prefix performed best, and in some cases the addition of the documentation string (docstring) improved the performance. Therefore, for the purposes of our study, we used 3 prompt variations, similar to TOGLL's evaluation with an extra addition:

\begin{itemize}
    \item method under test (mut) + test prefix
    \item docstring + mut + test prefix
    \item entire class under test (cut) + docstring + mut + test prefix

    To our knowledge, when evaluating test oracle generation techniques, there has never been used a prompt that includes the whole class under test. However, when generating the entire test suite, several studies provided the whole class in the instruction prompt. 

    In this study, it seemed relevant to include the entire class under test, in order to find out whether the LLMs are influenced by the implemented code. 
\end{itemize}

Finally, it is important to use the proper instruction to ensure good results. We designed two prompt instructions, one for each task, following Chattester's \cite{yuan2023no} approach. Similarly, both instructions begin with a role-playing text ("You are a professional who writes Java methods...") which serves as an optimization technique \cite{dong2023self}. Secondly, we include to the prompt its task which specifies that the LLM needs to answer based on the given code data. Finally, the last sentence specifies to the model the desired format of its answer. For the first case the answer is always binary (e.g. True/False, Correct/Incorrect), while for the second case the answer is typically 4-5 assertion oracles. Figure \ref{fig:prompt_structures} illustrates the structure of the prompt given to the LLM. Depending on the prompt variation, the applicable parts are filled. 

\begin{figure*}[t]
\centering

\begin{subfigure}[t]{0.48\textwidth}
\centering
\begin{tcolorbox}[
    enhanced,
    colback=black!3,
    colframe=black!40,
    boxrule=0.5pt,
    arc=2pt,
    width=\linewidth,
    height=6.5cm,
    valign=top,
    fontupper=\ttfamily\footnotesize
]

\textbf{[Class Under Test]} \\
Complete class under test \\(included only in Prompt Variation 3)

\vspace{0.4em}

\textbf{[Method Under Test]} \\
Method under test. If applicable, the docstring is prepended.

\vspace{0.4em}

\textbf{[Test Prefix]} \\
Test prefix without the oracle assertion

\vspace{0.4em}

\textbf{[Instruction]} \\
"You are a professional who writes Java test methods. Given the previous data, is the following assertion correct? Answer with only one word"

\vspace{0.4em}

\textbf{[Assertion]} \\
Assertion to be classified

\end{tcolorbox}

\caption{Prompt structure for classification tasks.}
\label{fig:prompt_classification}
\end{subfigure}
\hfill
\begin{subfigure}[t]{0.48\textwidth}
\centering
\begin{tcolorbox}[
    enhanced,
    colback=black!3,
    colframe=black!40,
    boxrule=0.5pt,
    arc=2pt,
    width=\linewidth,
    height=6.5cm,
    valign=top,
    fontupper=\ttfamily\footnotesize
]

\textbf{[Class Under Test]} \\
Complete class under test \\(included only in Prompt Variation 3)

\vspace{0.4em}

\textbf{[Method Under Test]} \\
Method under test. If applicable, the docstring is prepended.

\vspace{0.4em}

\textbf{[Test Prefix]} \\
Test prefix without the oracle assertion

\vspace{0.4em}

\textbf{[Instruction]} \\
"You are a professional who writes Java test methods in JUnit4 and Java 8. Given the previous data, generate 5 possible assertions. Answer with only 5 assertions"

\end{tcolorbox}

\caption{Prompt structure for generation tasks.}
\label{fig:prompt_generation}
\end{subfigure}

\caption{Prompt structures used for the classification and generation tasks.}
\label{fig:prompt_structures}

\end{figure*}
\subsection{Datasets}
\label{sec:datasets}

To conduct this experiment, we used two existing artifacts that contain both automated test suites and developer written tests. However, to cover all the aforementioned cases, the different prompt variations and the different tasks (classification or generation) we had to create a new dataset for each research question based on the two artifacts. Below follows a description of the two datasets and the final dataset created for the purposes of this experiment. 

\subsubsection{Open source Java repositories}

Firstly introduced for the (re-)evaluation of TOGA \cite{10.1145/3611643.3616265}, the used artifact contains 8 repositories that were used in Evosuite's benchmark and 17 repositories from the Apache Commons packages. In the same study, they generated test suites using Evosuite which were included in the artifact, and used for the purposes of our study as well. The same dataset was used to evaluate the performance of TOGA and TOGLL. Furthermore, combined with the fact that both aforementioned tools used Evosuite test prefixes to generate automated test suites, it seemed relevant to include the same Evosuite generated test prefixes as the ones used during their evaluation.

\subsubsection{Gitbug Java}

Gitbug java \cite{10555595} is a dataset introduced in 2024, which includes 199 bugs found in 55 open source repositories. The data are commits that include the bug, and the commit that fixed the bug. All the data are collected in commits occurred in 2023, and as indicated, there are no records in this dataset that have been \textit{seen} by the GPT model used in this experiment (GPT 3.5 Turbo).

\subsubsection{Construction of a new dataset(s)}

Although the first artifact was used to evaluate the two SOTA tools, its data structure lacked sufficient information to address all of our research questions. Since TOGA only requires the method under test, the docstring and the test prefix, the data found in the first artifact could not cover all the experiment cases we aimed for. Therefore, we followed their original procedure of dataset extraction, incorporating a few modifications to include additional information.

The artifact contained scripts that parse the test case generated by Evosuite, and extract the method under test, test prefix and assertion of each record. In our modification, we added the class under test, the package name and the test class name. The same procedure was done for the data found in Gitbug-Java. 
Furthermore, regarding the first artifact, we used $\mu$BERT with its default mutation operators to generate mutants representing cases where the code is wrong. Finally, we used the assertions that were wrong in RQ3 to construct the cases where the assertion is wrong. It is worth mentioning, that in our attempt to create an equally distributed dataset, we found out that the repository \textit{Async-http-client} did not contain many samples that include the documentation string. Hence, when generating the different cases and mutants, it was extremely difficult to find much data as only a handful of records contained information on all columns. Hence, the repository was dropped, leaving us only with 24 open source repositories from the first artifact. 

To summarize, in the end, we created a new dataset for each artifact found, and for each one of the four different cases. During the execution of the experiment, we set a maximum limit of using only 1000 records from each case in order to ensure that the results are based on an equally distributed dataset of 1000 records on each experiment case and each prompt variation. Furthermore, to ensure a fair comparison between the three prompt variations, any record that did not contain all the three fields of cut, docstring and mut, were excluded regardless of the prompt variation. This ensured that all prompt variations and experiment cases used the exact same records.

\subsection{Models}

We selected GPT-4o-mini~\cite{openai_gpt4o_mini} and DeepSeek V3~\cite{deepseek_v3_tech_report} for this experiment as they are representative of state-of-the-art large language models and are widely used in practice. These models also differ along important dimensions: GPT-4o-mini is a proprietary model, whereas DeepSeek V3 is open source, and both belong to different model families. In addition, we included GPT-3.5-turbo~\cite{brown2020languagemodelsfewshotlearners}, as it has been extensively used in prior work on automated test oracle generation and classification~\cite{yuan2023no, Yuan0DW00L24}, making it a suitable baseline for comparison. Finally, we incorporated Codestral~\cite{codestral_model_reference}, a code-specialized model, to evaluate whether our findings extend beyond general-purpose LLMs to models specifically optimized for programming tasks. 

To avoid randomness, we set the temperature as low as possible (temperature = 0.1). Furthermore, each dataset record was executed in an independent session, ensuring that no contextual information was shared across queries and that results were not influenced by prior interactions.

\subsection{Metrics}
In each one of the four analysed scenarios (correct/wrong code + correct/wrong assertion), we measure the LLM's performance in terms of its \emph{accuracy} in predicting the expected label for the given assertion. 
In the scenarios where the given assertion is the correct one (no matter if the code is correct or not), we expect that the model classify the assertion as correct. In the scenarios where the assertion is incorrect, we expect the model to predict it as incorrect. 
Typically the accuracy in a classification problem is calculated using the true positives, true negatives, false positives and false negatives obtained. We refer it in the paper as \textit{Overall accuracy}:

\begin{equation}
Accuracy = \frac{TP + TN}{TP + TN + FP + FN}
\end{equation}

However, because we break the data into the four analysed scenarios, in practice the problem becomes binary and contains only two metrics: true classification (hit) or misclassification (miss). 

Such accuracy is calculated as the number of hits divided by the total number of records. In other words, the following formula indicates the accuracy of the model for this research question.

\begin{equation}
Accuracy = \frac{Hits}{Hits + Misses}
\end{equation}

We measure the strength of the test oracle assertions in terms of the mutation scores they can achieve, i.e.the ratio of killed mutants among all the generated mutants (cf. Section~\ref{sec:mutation-testing}).

\begin{equation}
Mutation~Score = \frac{Killed~mutants}{Killed~mutants + Surviving~mutants}
\end{equation}

\section{Results}

\subsection{RQ1: (How accurately) Can LLMs find the actual or the expected oracle in a classification problem?}

\begin{figure*}[b]
    \centering
    \includegraphics[width=1\linewidth]{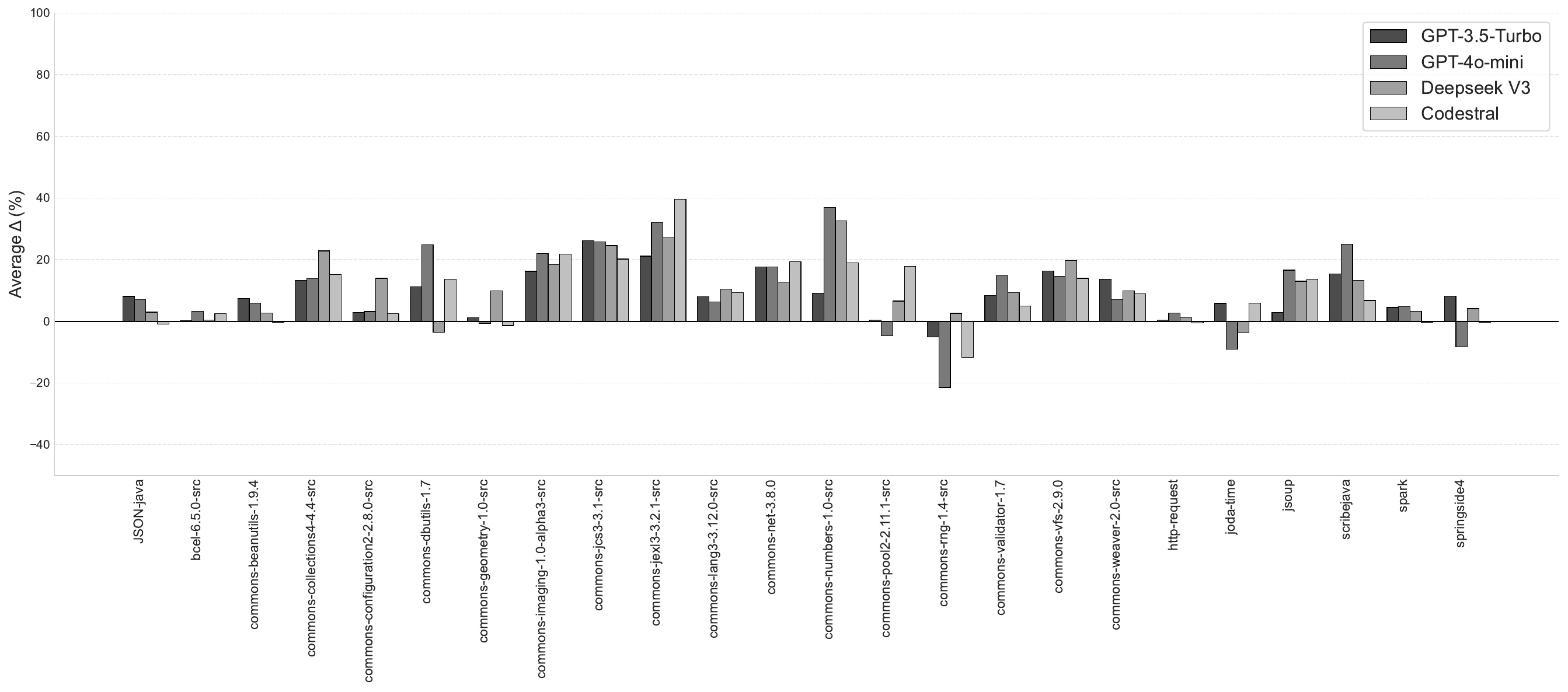}
    \caption{RQ1: Average difference ($\Delta$) on the accuracy between experiment cases 1 and 2 for each repository. In both cases, the assertion is the same and correct, but the implementation given is different. The figure shows that, for the majority of repositories, there is a notable difference between the two configurations, suggesting that the LLMs are influenced by the actual implementation rather than the expected behaviour.}
    \label{fig:rq1_diff_results}
\end{figure*}

Table \ref{table:avg_results} presents the average accuracy across all repositories for each prompt variation, under the four studied scenarios. For instance, ``CC + CA'' refers to the scenario in which both the code and the test oracle assertion are correct. 
Moreover, Figure \ref{fig:rq1_diff_results} illustrates the difference of the accuracy between the first two scenarios, where in both cases the given test assertion is correct, but in the first one the code is correct and in the second one is incorrect (buggy). 
Overall, when the assertion is correct, all LLMs achieve a higher classification accuracy when fed with the correct code, rather than when the given code is buggy. 
This indicates that, correct assertions are more likely to be misclassified by the LLMs when the given code is slightly mutated (buggy). 

\begin{finding}
LLMs' test oracle classification accuracy considerably drops in the presence of buggy code, suggesting that its predictions are derived towards the actual implementation rather than the desired one. 
\end{finding}

Another notable observation is that when incorrect assertions are provided (scenarios CC + WA and WC + WA in Table \ref{table:avg_results}), the accuracy of the LLMs increases. In particular, the highest performance is observed when both the assertion and the code are incorrect. Combined with the results observed in the other two cases (correct assertion given), these findings suggest that the models have more confidence in rejecting assertions when uncertain. 

The only exception to this pattern is the code-oriented model Codestral. While Codestral also exhibits improved accuracy when both the code and the assertion are incorrect (WC + WA), it performs better in identifying correct assertions when the code itself is correct (CC + CA) but performs poorly when the assertion is incorrect (CC + WA). This indicates that Codestral is less inclined to reject assertions and instead shows a stronger tendency to accept them in the presence of correct code.

\begin{finding}
General-purpose LLMs tend to favor rejecting assertions rather than accepting them. In contrast, the code-oriented model Codestral exhibits the opposite behavior, showing a greater tendency to accept assertions. Across all models, however, higher accuracy is observed when both the code and the assertion are incorrect.
\end{finding}

Finally, it is important to examine the differences across prompt variations. Overall, we observe that model performance tends to improve when the code of the class under test is provided; however, this trend is not entirely consistent across all experimental settings. The last column of Table~\ref{table:avg_results} summarizes the overall accuracy achieved under each prompt variation. Across all models, the highest classification accuracy is observed for the final prompt variation, which includes both the class under test and the docstring. 

\begin{table*}[ht]
\centering
\caption{RQ1: Accuracy for each experiment case and each prompt variation. For reference, the overall accuracy is reported which is a combination of all four experiment cases \\ (CC = Correct Code, WC = Wrong Code, CA = Correct Assertion, WA = Wrong Assertion)}
\label{table:avg_results}
\renewcommand{\arraystretch}{1.2}
\setlength{\tabcolsep}{8pt}
\resizebox{\textwidth}{!}{
\begin{tabular}{c | r | c c c c | c}
\toprule
\textbf{Model} & \textbf{Prompt Variation} 
& \textbf{CC + CA} & \textbf{WC + CA} & \textbf{CC + WA} & \textbf{WC + WA} & \textbf{Overall} \\
\midrule
\multirow{3}{*}{Deepseek V3} 
 & MUT + TP & 43.02\% & 35.72\% & 74.17\% & 96.53\% & 62.07\% \\
 & Docstring + MUT + TP & 49.47\% & 41.17\% & 69.70\% & 95.09\% & 63.59\% \\
 & CUT + Docstring + MUT + TP & 55.17\% & 39.02\% & 81.05\% & 96.10\% & 67.54\% \\
\hline
\multirow{3}{*}{GPT-3.5-turbo} 
 & MUT + TP & 39.18\% & 31.20\% & 56.47\% & 84.60\% & 52.47\% \\
 & Docstring + MUT + TP & 44.76\% & 36.42\% & 52.19\% & 80.39\% & 53.08\% \\
 & CUT + Docstring + MUT + TP & 44.31\% & 37.18\% & 62.47\% & 84.32\% & 56.87\% \\
\hline
\multirow{3}{*}{GPT-4o-mini} 
 & MUT + TP & 51.15\% & 44.95\% & 57.90\% & 83.49\% & 59.26\% \\
 & Docstring + MUT + TP & 54.46\% & 47.36\% & 56.21\% & 77.01\% & 58.48\% \\
 & CUT + Docstring + MUT + TP & 60.33\% & 46.58\% & 66.30\% & 77.52\% & 62.53\% \\
\hline
\multirow{3}{*}{Codestral} 
 & MUT + TP & 70.84\% & 62.34\% & 36.12\% & 80.59\% & 62.33\% \\
 & Docstring + MUT + TP & 71.93\% & 63.01\% & 39.02\% & 77.92\% & 62.84\% \\
 & CUT + Docstring + MUT + TP & 80.55\% & 72.02\% & 48.42\% & 76.95\% & 69.42\% \\
\bottomrule
\end{tabular}
}
\end{table*}

\subsection{RQ2: What can influence the finding of an expected oracle instead of an actual oracle?}

To answer to this question we analysed 1000 test prefixes taken from the Gitbug-Java dataset.  
Table \ref{tab:rq2_results} reflects how the LLMs' accuracy is affected while more noise (out of distribution test and variable names) are injected in the test prefixes. 
Notice that the models obtain the highest accuracy, when the original test code is used. 
However, while more modifications are included, the more the accuracy of the model drops. 

Particularly, the second prompt variation resulted in the best performing in all the cases. 
Notably, the prediction accuracy between the original code given and the most noisy one (similarly to Evosuite, where test and variables have meaningless names), the results dropped by 16,10\% for the first prompt variation, 9,80\% for the second prompt variation, and 2,70\% on the third prompt variation. 

When only the test name was modified to resemble Evosuite's naming convention, it can already be seen that the results dropped except when the third prompt variation was used. For the first prompt variation the accuracy dropped by 6,20\%, for the second prompt variation the accuracy dropped by 5,90\% and for the third prompt variation the accuracy increased by just 0,60\%.

\begin{table*}[ht]
\centering
\caption{RQ2: Accuracy (\%) of assertion classification, after test modifications, grouped by prompt variation and LLM used}
\label{tab:rq2_results}
\renewcommand{\arraystretch}{1.2}
\setlength{\tabcolsep}{8pt}
\resizebox{\textwidth}{!}{
\begin{tabular}{c | r | c c l}
\toprule
\textbf{Model} & \textbf{Prompt Variation} 
& \textbf{Original} & \textbf{Modified test names} & \textbf{Modified test names and vars.} \\
\midrule
\multirow{3}{*}{DeepSeek V3} 
 & MUT + TP & 44.70\% & 45.20\% & 40.30\%\hspace{1mm}(-4.40) \\
 & Docstring + MUT + TP & 54.10\% & 54.50\% & 47.90\%\hspace{1mm}(-6.20) \\
 & CUT + Docstring + MUT + TP & 49.90\% & 49.80\% & 39.70\%\hspace{1mm}(-10.20) \\
\hline
\multirow{3}{*}{GPT-3.5-Turbo} 
 & MUT + TP & 44.00\% & 38.70\% & 28.70\%\hspace{1mm}(-15.30) \\
 & Docstring + MUT + TP & 53.40\% & 47.40\% & 43.40\%\hspace{1mm}(-10.00) \\
 & CUT + Docstring + MUT + TP & 43.50\% & 44.00\% & 41.10\%\hspace{1mm}(-2.40) \\
\hline
\multirow{3}{*}{GPT-4o-mini} 
 & MUT + TP & 55.20\% & 53.40\% & 42.60\%\hspace{1mm}(-12.60) \\
 & Docstring + MUT + TP & 59.50\% & 56.00\% & 44.70\%\hspace{1mm}(-14.80) \\
 & CUT + Docstring + MUT + TP & 62.50\% & 58.20\% & 47.10\%\hspace{1mm}(-15.40) \\
\hline
\multirow{3}{*}{Codestral} 
 & MUT + TP & 86.20\% & 85.90\% & 82.40\%\hspace{1mm}(-3.80) \\
 & Docstring + MUT + TP & 89.00\% & 87.10\% & 82.40\%\hspace{1mm}(-6.60) \\
 & CUT + Docstring + MUT + TP & 88.60\% & 89.20\% & 84.90\%\hspace{1mm}(-3.70) \\
\bottomrule
\end{tabular}
}
\end{table*}

\begin{finding}
Descriptive test and variable names improve LLMs' predictions. Under "anonymized" test names and variables, the accuracy of GPT-3.5-turbo, GPT-4o-mini, DeepSeek V3 and Codestral dropped up to15.30\%, 15.40\%, 10.20\% and 6.60\% respectively. 
\end{finding}

\subsection{RQ3: (How accurately) Can LLMs generate a test oracle in a text generation problem?}

\textbf{Acc@1 - Generating single assertion:}
Table~\ref{tab:acc1_generation_results} reports the average compilation rate and passing rate (i.e., accuracy) of the generated test oracles when the LLM is instructed to produce a single assertion. Since the code under test is always correct, passing assertions directly reflect the model’s accuracy to generate valid oracles. 

Consistent with previous research questions, the third prompt variation (which includes the class under test) generally yields the best performance across both compilation and passing rates. The only exception is observed for DeepSeek V3, where the second prompt variation achieves a slightly higher compilation rate.

A test in our experiments can fail for two distinct reasons: either due to syntactic invalidity or semantic mismatch. A closer examination of the results in the table provides further insight into where LLMs struggle. For GPT-3.5-turbo, both compilation and passing rates are relatively low, indicating difficulties at both the syntactic and semantic levels. For the more recent models, however, both rates improve significantly, with the models generating syntactically valid oracles in nearly 9 out of 10 cases. This observation suggests that modern LLMs are generally capable of producing well-formed code, but still face challenges in correctly capturing the intended behaviour of the tests.

GPT-4o-mini generated 1.74, 2.49, and 2.69 times more non-passing oracles than non-compiling ones across the three prompt variations. Similarly, DeepSeek V3 produced 1.92 and 2.92 times more non-passing assertions for the first two prompt variations, while exhibiting nearly equal numbers of non-passing and non-compiling oracles for the third variation. Codestral shows the largest disparity between the two failure modes, generating 2.58, 2.81, and 4.79 times more non-passing tests than syntactically invalid ones across the three prompt variations. Lastly, GPT-3.5-turbo is the only model for which we observe a higher number of syntactically invalid oracles compared to semantically invalid ones. Under the first prompt variation, it produced 0.93 times as many non-passing tests as non-compiling ones, indicating a slight dominance of syntactic failures. However, this trend shifts in the subsequent prompt variations, where the model generates 1.16 and 1.78 times more non-passing tests than non-compiling tests, respectively.

\begin{finding}
 Modern LLMs generate syntactically valid oracles in nearly 9 out of 10 cases, with failures predominantly arising from semantic mismatches rather than compilation errors. This pattern is consistent across GPT-4o-mini, DeepSeek V3, and Codestral, indicating that semantic correctness remains the primary challenge in oracle generation, whereas GPT-3.5-turbo shows comparatively weaker performance on both syntactic and semantic fronts.
\end{finding}

\begin{table*}[bt]
\centering
\caption{RQ3: Acc@1 generation task results. The average compilation and passing rates are shown for each model and prompt variation. In this experiment, the code under test was always correct, and the model had to generate one assertion. }
\label{tab:acc1_generation_results}
\resizebox{\textwidth}{!}{%
\begin{tabular}{r|rr|rr|rr|rr}
\toprule
\multirow{2}{*}{Prompt Variation} 
 & \multicolumn{2}{c|}{\textbf{GPT-3.5-turbo}} 
 & \multicolumn{2}{c|}{\textbf{GPT-4o-mini}} 
 & \multicolumn{2}{c|}{\textbf{Codestral}} 
 & \multicolumn{2}{c}{\textbf{DeepSeek V3}} \\
\cmidrule(lr){2-3} \cmidrule(lr){4-5} \cmidrule(lr){6-7} \cmidrule(lr){8-9}
 & Comp.\ Rate & Pass.\ Rate 
 & Comp.\ Rate & Pass.\ Rate 
 & Comp.\ Rate & Pass.\ Rate 
 & Comp.\ Rate & Pass.\ Rate \\
\midrule
MUT + TP 
& 73.35\% & 48.66\% 
& 87.35\% & 65.27\% 
& 89.29\% & 61.60\% 
& 87.96\% & 64.88\% \\
Docstring + MUT + TP 
& 77.12\% & 50.53\% 
& 90.88\% & 68.17\% 
& 90.11\% & 62.31\% 
& 92.16\% & 69.25\% \\
CUT + Docstring + MUT + TP 
& 86.46\% & 62.36\% 
& 93.76\% & 76.98\% 
& 94.68\% & 69.24\% 
& 87.94\% & 75.90\% \\
\bottomrule
\end{tabular}
}
\end{table*}

Although classification cannot directly compare to the generation task, we consider two comparisons to estimate whether it is more effective to prompt LLMs to classify or to generate assertions:
\begin{enumerate}
    \item Overall classification accuracy versus generation accuracy.
    \item Classification accuracy under correct code (CC + CA and CC + WA) versus generation accuracy. Since the generation task was conducted on correct code, it might make more sense to compare the two tasks on the same records.
\end{enumerate}

Table~\ref{table:generation_versus_classification} compares generation accuracy (i.e., passing rate) with both the overall classification accuracy from RQ1 and the classification accuracy restricted to correct-code cases. Across both comparisons, all general-purpose models achieve higher accuracy in the generation setting, suggesting that LLMs are more effective when tasked with generating test oracles than when classifying them.

Codestral, the code-oriented model, exhibits slightly higher performance in the classification task although the differences are minimal. Overall, the generation task consistently yields higher accuracy than classification under correct-code-only samples.

Nevertheless, these results do not imply that generated oracles are inherently more effective at detecting faults. Rather, they indicate that, given the same test prefixes, LLMs are more likely to produce a passing oracle than to correctly classify an existing one.

\begin{finding}
LLMs are more effective at generating test oracles than at classifying them. When provided with regression test prefixes, all models produce a higher number of passing assertions in the generation setting than in the classification setting.
\end{finding}

\begin{table*}[bt]
\centering
\caption{\textbf{RQ3: Comparison between Generation and Classification.} The table reports the classification accuracy over the full dataset, as well as the classification accuracy restricted to cases where the code under test is correct, alongside the accuracy of the generation task.}
\label{table:generation_versus_classification}
\renewcommand{\arraystretch}{1.2}
\setlength{\tabcolsep}{8pt}
\resizebox{\textwidth}{!}{
\begin{tabular}{c | r | c  c  c}
\toprule
\textbf{Model} & \textbf{Prompt Variation} 
& \textbf{Generation accuracy} & \textbf{Classification accuracy} & \textbf{Classification (Correct-code only) accuracy} \\
\midrule
\multirow{3}{*}{Deepseek V3} 
 & MUT + TP & \textbf{64.88\%} & 62.07\% & 58.72\% \\
 & Docstring + MUT + TP & \textbf{69.25\%} & 63.59\% & 59.63\% \\
 & CUT + Docstring + MUT + TP & \textbf{75.90\%} & 67.54\% & 68.21\% \\
\hline
\multirow{3}{*}{GPT-3.5-turbo} 
 & MUT + TP & \textbf{48.66\%} & 52.47\% & 47.69\% \\
 & Docstring + MUT + TP & 50.53\% & \textbf{53.08\%} & 48.44\% \\
 & CUT + Docstring + MUT + TP & \textbf{62.36\%} & 56.87\% & 53.56\% \\
\hline
\multirow{3}{*}{GPT-4o-mini} 
 & MUT + TP & \textbf{65.27\%} & 59.26\% & 54.72\% \\
 & Docstring + MUT + TP & \textbf{68.17\%} & 58.48\% & 55.10\% \\
 & CUT + Docstring + MUT + TP & \textbf{76.98\%} & 62.53\% & 63.39\% \\
\hline
\multirow{3}{*}{Codestral} 
 & MUT + TP & 61.60\% & \textbf{62.33\%} & 53.34\% \\
 & Docstring + MUT + TP & 62.31\% & \textbf{62.84\%} & 55.34\% \\
 & CUT + Docstring + MUT + TP & 69.24\% & \textbf{69.42\%} & 64.36\% \\
\bottomrule
\end{tabular}
}
\end{table*}

\begin{figure}[b]
\centering

\begin{subfigure}{0.49\linewidth}
    \centering
    \includegraphics[width=\linewidth]{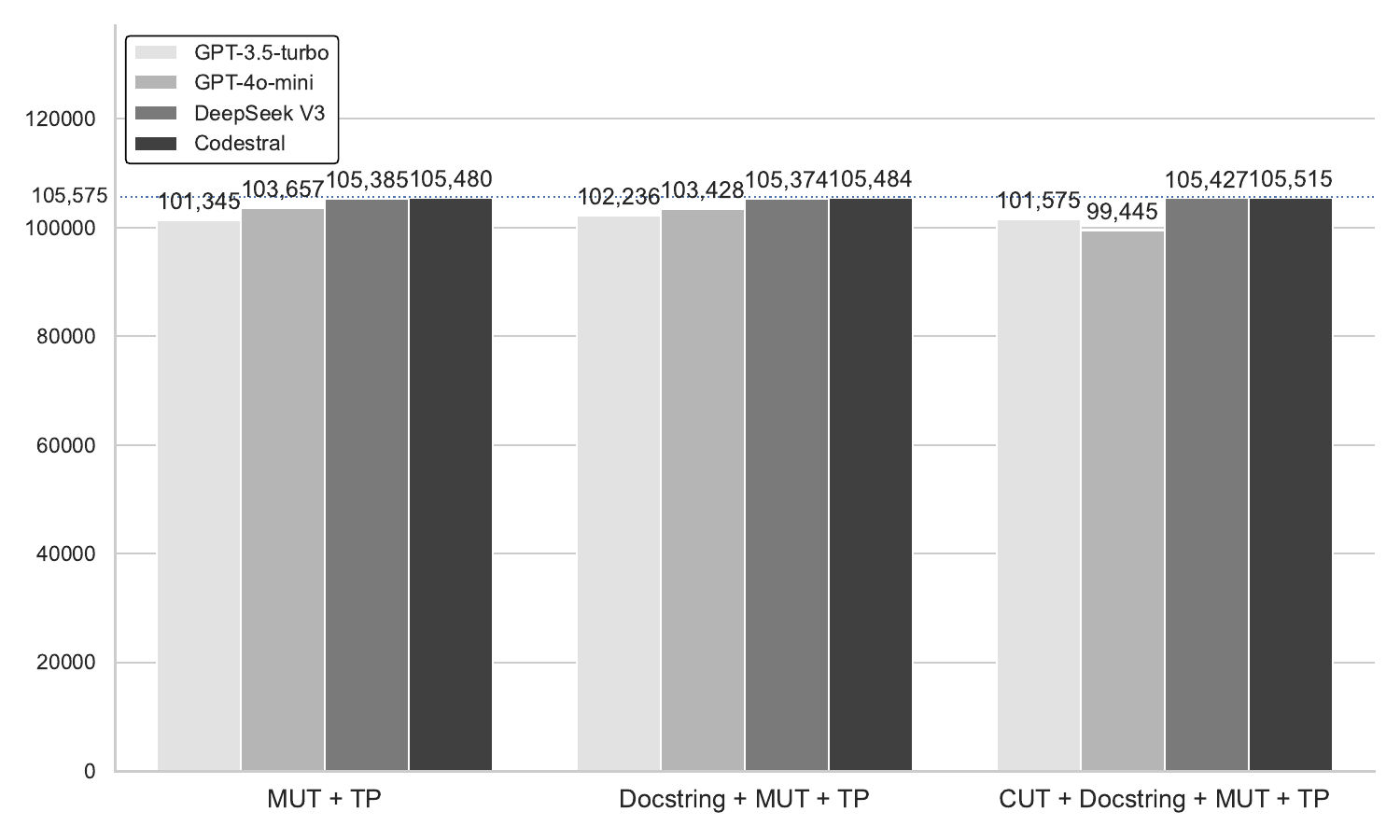}
    \caption{Number of generated assertions}
    \label{fig:plot_a}
\end{subfigure}
\hfill
\begin{subfigure}{0.49\linewidth}
    \centering
    \includegraphics[width=\linewidth]{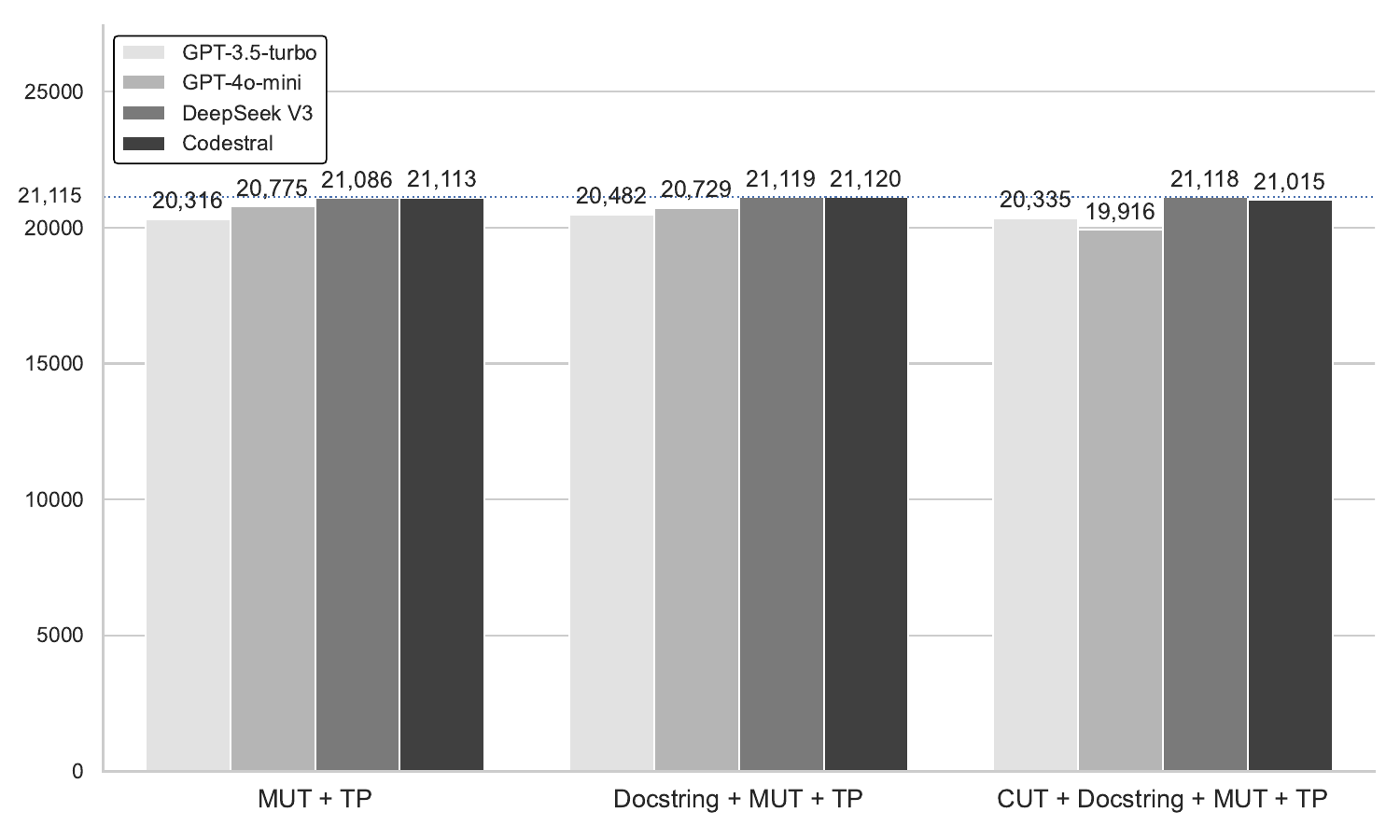}
    \caption{Number of generated tests}
    \label{fig:plot_b}
\end{subfigure}

\caption{Acc@5: Number of generated assertions (left) and the corresponding number of generated tests (right) across prompt variations and models. The maximum possible number of assertions is 105,575 and the maximum number of tests is 21,115. None of the models reach these upper bounds. In several cases, fewer than five assertions were generated per task, despite instructions to generate five.}
\label{fig:acc5_meta}
\end{figure}

\textbf{Acc@5 - Generating five assertions:} Similarly to Acc@1 setup, we take a maximum of 1000 test prefixes for each project in TOGLL~\cite{hossain2024togllcorrectstrongtest} dataset, and instruct the LLM to generate five test oracle assertions to complete each test prefix. 
In few cases the model generated less assertions than five (typically four), and in very few occasions the models did not produce any assertion and thus were discarded. Figure~\ref{fig:acc5_meta} reports the total number of generated assertions and test cases. While the theoretical maximum is 105,575 assertions and 21,115 tests, none of the models reached these limits. Among the evaluated models, Codestral and DeepSeek V3 generated the highest number of assertions, coming closest to the theoretical maximum.

Table~\ref{table:rq3_results} presents the results for each prompt variation in terms of accuracy when the models are instructed to generate five assertions. To enable a more comprehensive and fair evaluation, we consider two complementary metrics:

\begin{enumerate}
    \item \textbf{Accuracy of all generated assertions:} This metric measures the proportion of all generated assertions that pass, reflecting the overall correctness of the generated outputs.
    
    \item \textbf{Acc@5:} Following prior work~\cite{10804561}, this metric evaluates accuracy per test, by checking whether at least one of the generated assertions (out of five) passes.
\end{enumerate}

\begin{table}[t]
\centering
\caption{RQ3: Results: Table presents the average accuracy between all generated assertions per repository. Additionally Acc@5 is reported, meaning the number of cases where at least 1 out of 5 generated assertions passed.}
\label{table:rq3_results}
\begin{tabular}{r | c c c c}
\toprule
Prompt Variation 
& GPT-3.5-Turbo 
& GPT-4o-mini 
& Codestral 
& DeepSeek V3 \\
\midrule
\multicolumn{5}{c}{\textbf{Average Accuracy: All generated assertions}} \\
\midrule
MUT + TP
& 58.95\% & 62.16\% & 57.82\% & 58.78\% \\
Docstring + MUT + TP
& 57.47\% & 64.09\% & 61.04\% & 64.68\% \\
CUT + Docstring + MUT + TP
& 60.01\% & 71.85\% & 69.14\% & 76.26\% \\
\midrule
\multicolumn{5}{c}{\textbf{Acc@5: Accuracy when at least 1 assertion is passing}} \\
\midrule
MUT + TP
& 93.76\% & 95.44\% & 92.73\% & 84.05\% \\
Docstring + MUT + TP
& 91.31\% & 97.06\% & 94.49\% & 89.21\% \\
CUT + Docstring + MUT + TP
& 89.34\% & 97.34\% & 92.38\% & 92.86\% \\
\bottomrule
\end{tabular}
\end{table}

When analysing the \textit{Acc@5} metric, we can observe that in around 90\% of the cases, the LLMs managed to produce a test assertion that passes the test. This result is expected, as generating multiple assertions increases the likelihood of producing a correct one. Consequently, higher accuracy under Acc@5 compared to Acc@1 is not surprising.

More notably, the accuracy of all generated assertions does not decrease proportionally when moving from Acc@1 to Acc@5; in some cases, it even improves. It is worth noting that, across all models, the highest generation accuracy is typically observed in one of the Acc@1 prompt variations. However, when models are instructed to generate five assertions, they produce a larger set of candidates, thereby increasing the likelihood of generating a correct assertion.

This represents a favorable trade-off: although individual assertion quality does not always improve, the availability of multiple assertions increases the chances of identifying a correct oracle. As a result, more test prefixes can be successfully completed, leading to a higher number of usable test cases under the Acc@5 setting.

\begin{finding}
Generating multiple assertions (Acc@5) increases the likelihood of producing a correct oracle per test. Perhaps surprisingly, this does not necessarily lead to a reduction in overall assertion quality. Under the Acc@5 setting, \textbf{GPT-3.5-turbo}, \textbf{GPT-4o-mini}, \textbf{Codestral}, and \textbf{DeepSeek V3} generate at least one passing assertion in up to \textbf{93.76\%}, \textbf{97.34\%}, \textbf{94.49\%}, and \textbf{92.86\%} of the cases, respectively. These results are substantially higher than those observed under the Acc@1 setting.
\end{finding}

When it comes to the prompt variations, we observe something rather interesting. For accuracy of all generated assertions, the third prompt variation is the one which consistently performs the best. Interestingly, this does not necessarily reflect also their respective Acc@5 results. Although GPT-4o-mini and DeepSeek V3 have the same trend, GPT-3.5-Turbo has better Acc@5 under the first prompt variation and Codestral has better Acc@5 under the second prompt variation.

\subsection{RQ4: How strong are the oracles generated by the LLM?}

\begin{table}[b]
\centering
\caption{\textbf{ACC@5: Average mutation score (\%) across models and prompt variations:} For each passing generated oracle, we generated a test case. The table shows the mutation score obtained with test cases of the passing LLM-generated oracles, and the mutation score obtained when the same test cases were created with their respective Evosuite oracles.}
\label{table:rq4_avg_results}
\resizebox{\textwidth}{!}{
\begin{tabular}{l r cc}
\toprule
Model & Prompt Variation & Model's Mutation score & EvoSuite's mutation score \\
\midrule
\multirow{3}{*}{GPT-3.5-Turbo}
& MUT + TP & \textbf{19.92\%} & 17.56\% \\
& Docstring + MUT + TP & \textbf{18.01\%} & 17.29\% \\
& CUT + Docstring + MUT + TP & \textbf{19.37\%} & 17.11\% \\
\midrule
\multirow{3}{*}{GPT-4o-mini}
& MUT + TP & 18.76\% & \textbf{19.84\%} \\
& Docstring + MUT + TP & \textbf{18.76\%} & 18.74\% \\
& CUT + Docstring + MUT + TP & \textbf{19.77\%} & 19.16\% \\
\midrule
\multirow{3}{*}{Codestral}
& MUT + TP & \textbf{20.58\%} & 19.61\% \\
& Docstring + MUT + TP & \textbf{19.82\%} & 18.69\% \\
& CUT + Docstring + MUT + TP & \textbf{20.62\%} & 19.51\% \\
\midrule
\multirow{3}{*}{DeepSeek V3}
& MUT + TP & \textbf{18.70\%} & 18.36\% \\
& Docstring + MUT + TP & \textbf{18.55\%} & 18.43\% \\
& CUT + Docstring + MUT + TP & \textbf{19.57\%} & 18.75\% \\
\bottomrule
\end{tabular}
}
\end{table}

We compute and compare the mutation score of the test suites generated by including the LLM and Evosuite generated test oracle assertions, respectively. 
Notice that, for each prompt variation and project, we only consider the test cases for which the LLM generated at least one assertion that passes the tests on the original code. Afterwards, we run PiTest to calculate the mutation scores and summarise. We chose the results of Acc@5 for such an evaluation, as the likelihood of LLM to generate a generic/non-relevant oracle is higher when prompted to generate five assertions. 

Table~\ref{table:rq4_avg_results} shows that the mutation score achieved by test suites generated with LLM-generated oracles is generally higher than that of the corresponding EvoSuite-generated test suites, with only one exception. Specifically, GPT-4o-mini under the first prompt variation yields slightly lower mutation score compared to EvoSuite.

These findings suggest that as a minimum, the LLM-generated oracles are not overly generic or less relevant than the Evosuite-generated ones with respect to the class under test. To the best of our knowledge, there is no established metric for directly quantifying oracle relevance or for comparing different oracle generation approaches. Nevertheless, the results of this study suggest that LLMs are able to generate oracles that lead to higher mutation score than traditional approaches, which may provide stronger fault-detection capabilities.

\begin{finding}
LLMs can generate stronger oracles than traditional approaches, leading to higher fault detection capabilities. This suggest that LLMs can be used to complement and improve test case generation.
\end{finding}

\section{Discussion}

\subsection{Oracle relevance in generation and classification tasks}

RQ4 was designed to assess the extent to which oracles generated by LLMs are \textit{relevant}. In principle, generating test oracles can be a straightforward task, as one can execute the program under test and derive assertions that validate its current behaviour. However, such oracles may only capture the \textit{implemented} behaviour rather than the \textit{intended} one. To effectively detect faults, oracles must reflect the expected behaviour of the system.

Assessing oracle relevance remains challenging, as no established metric exists for directly quantifying it. In this work, we approximate oracle relevance through mutation testing. The results indicate that LLM-generated oracles are not overly generic and are, on average, at least as effective as those produced by traditional approaches, if not better. However, mutation testing is not entirely accurate, as in theory a model can generate more assertions that will increase its mutation score. This does not necessarily indicate that the generated oracles are more relevant though. 

A closer examination of the classification results further suggests that LLMs exhibit useful capabilities in this context. As shown in RQ1, the models achieve relatively high accuracy in identifying incorrect (externally generated) assertions. As seen in RQ1, the LLM has a high accuracy in classifying (externally generated) incorrect assertions as negative.  In a context in which assertions are first generated and then filtered out, such in TOGA and TOGLL, LLMs are good candidates to discard the invalid ones.

We manually investigated some assertions in which the LLM  misclassified as incorrect (i.e. these were correct but were predicted as invalid). 
We asked the model to justify \textit{why} did it answer that the assertion was incorrect. 
Below we report the justification provided by the LLM in two examples:

\begin{tcolorbox}[colback=gray!10, colframe=gray!50,boxrule=0.4pt,arc=1pt,left=6pt,right=6pt,top=4pt,bottom=4pt,fontupper=\small,before skip=8pt,after skip=8pt,enhanced]
\textbf{Example 1:} \textit{"The assertion is incorrect because it does not validate any specific behaviour of the averageOptionVolumeAbove() method. It simply checks if the result is 1, which doesn't provide any meaningful validation of the method's functionality."}
\end{tcolorbox}

\begin{tcolorbox}[colback=gray!10, colframe=gray!50,boxrule=0.4pt,arc=1pt,left=6pt,right=6pt,top=4pt,bottom=4pt,fontupper=\small,before skip=8pt,after skip=8pt,enhanced]
\textbf{Example 2:} \textit{"The assertion is incorrect because it's comparing the expected value Integer.MAX\_VALUE with the actual value returned by scannerSubscription0.averageOptionVolumeAbove(), which is also Integer.MAX\_VALUE. This assertion will pass, but it doesn't effectively verify the behaviour of the method under test. The method averageOptionVolumeAbove() should be returning the default value Integer.MAX\_VALUE, which is what the assertion is comparing against. However, to effectively test the method, you should set a different value for averageOptionVolumeAbove() and then verify if the method returns that value correctly."}
\end{tcolorbox}

The model often replied that the assertion was correct but too vague. In other words, the LLM rejected some correct assertions because they did not seem relevant to the test's behaviour. 
We believe that this limitation can be mitigated by incorporating a second LLM-as-a-Judge, to increase the confidence that the assertion should be discarded. At the same time, this perhaps reveals a possible future direction, where LLMs are being used with respect to the oracle relevance problem. 

To conclude, while this paper investigates whether LLMs can classify or generate fault-detecting oracles, its findings may point towards a promising future direction for detecting oracle relevance.

\subsection{Prone to capture the actual behaviour, not the expected one}

Our experiments showed that the LLM's accuracy to correctly classify a correct assertion as positive, drops when the given code is buggy. 
This suggest that the LLM is prone to follow the actual implementation to classify the test oracle rather than the expected behaviour. 
To gain further insights, we asked the model to justify its decision. Below we provide one example:

\begin{tcolorbox}[colback=gray!10, colframe=gray!50,boxrule=0.4pt,arc=1pt,left=6pt,right=6pt,top=4pt,bottom=4pt,fontupper=\small,before skip=8pt,after skip=8pt,enhanced]
\textbf{Example 3:} \textit{"The assertion is incorrect because the method averageOptionVolumeAbove() in the ScannerSubscription class is expected to return -2, not Integer.MAX\_VALUE."}
\end{tcolorbox}

Precisely, the LLM justification followed the information found on the actual code implementation and based its assertion on conclusions that can be derived from the implementation. Hence, although multiple studies use LLMs as if they would extract the expected oracle, our empirical results suggest that the LLM follows the actual code execution to find the test oracle. 

\subsection{Reasoning models and their identical behaviour}

To investigate whether newer models with reasoning capabilities exhibit different behaviour from that observed in our main experiments, we repeated the experiments corresponding to the first research question. Specifically, we examined whether reasoning models remain inclined towards the actual behaviour of the implementation rather than its expected behaviour. Table \ref{tab:reasoning_models} describes the results obtains under GPT-5-mini.

For GPT-5-mini, we observe that our main findings persist. When evaluated on the correct implementation, the model achieves high accuracy, reaching 92.45\% when the class under test is included in the prompt. However, when the implementation is mutated (wrong code condition), accuracy decreases by 18.65\% to 27.06\%, depending on the prompt variation. Interestingly, the third prompt variation, which achieves the highest accuracy under the correct implementation, also exhibits the largest decrease in accuracy when evaluated against the mutated implementation.

\begin{finding}
Similarly to non-reasoning models, reasoning models are influenced by the actual implementation rather than the expected behaviour. This tendency is consistently observed across all prompt variations.
\end{finding}

\begin{table}[t]
    \centering
    \caption{Replication of RQ1: Accuracy of GPT-5-mini (reasoning model) under all prompt variations.}
    \label{tab:reasoning_models}
    \begin{tabular}{rcc}
        \toprule
        \textbf{Prompt Variation} & \textbf{CC + CA} & \textbf{WC + CA} \\
        \midrule
        MUT + TP                    & 80.83\% & 61.76\% \\
        Docstring + MUT + TP        & 82.77\% & 64.12\% \\
        CUT + Docstring + MUT + TP  & 92.45\% & 65.39\% \\
        \bottomrule
    \end{tabular}
\end{table}

\subsection{Prompt Optimization and Design do not affect the performance}

In our main experiments, we did not employ any advanced prompt optimization techniques (e.g., Chain-of-Thought prompting). Although several prompt formulations were explored during experimentation, the core instruction text remained semantically equivalent across runs.

To assess whether prompt design influences how LLMs reason about assertions, specifically whether they rely on the actual implementation versus the expected specification, we replicated the experiments from RQ1 using alternative prompt strategies with \texttt{gpt-4o-mini}.

The first prompt optimization technique we used was to split the classification prompt into two steps. Instead of directly asking the LLM to classify the given assertion, we asked the LLM to summarise the behaviour of the method under test, similarly to ChatAssert's generation process~\cite{10804561}. For the second prompt variation, we introduced an explicit instruction warning the model that \emph{the code might be buggy}, aiming to discourage reliance on the observed code behavior.

While these prompt variations resulted in minor differences in overall accuracy, they did not alter the fundamental reasoning pattern of the model. In all configurations, the LLM predominantly grounded its classification decisions based on the actual implementation rather than the expected behavior. 

\subsection{Meaningful names impact the performance}

Recent studies suggest that standard test generation approaches, such as Evosuite, can be used to generate effective test prefixes (e.g. to cover all branches) and then use an AI-based solution (e.g. an LLM) to generate the test oracle. 
While this is a promising line of research, RQ2 provided concrete evidence that automatically generated test/variable names can considerably drop the LLM effectiveness. 
In particular, the model is much more precise to correctly classify correct assertions as positive when the test uses developer-like naming conventions. 
This finding is inline with recent studies~\cite{SallouDP2024} that show that LLM's performance are affected when evaluated on out-of-distribution inputs. 

In order to understand better why the model performs better on developer written tests, we asked the LLM to provide us with the justification on such predictions. 
While the code was still taken into consideration, the information provided by the test and variable names, such as \textit{shouldFail} alongside relevant exceptions in the test, provided valuable insights to the model to capture the test's expected behaviour. Therefore, the LLM could guess better what should be the expected test assertion, regardless the actual code implementation. 

This suggest that, if LLMs are meant to be used to classify/generate the test oracle, will be determinant to generate tests that follow developer-like naming conventions. 

\subsection{Documentation and Alternative sources of expected behaviour}

The results of this study indicate that the quality and richness of input information influence the performance of LLMs. In particular, meaningful identifiers (e.g., method and variable names) and additional contextual information such as documentation and class-level code, might improve the quality of generated outputs. This suggests that LLM performance is closely tied to the extent to which the provided input captures the intended or expected behaviour of the system.

More specifically, sources such as documentation are more likely to explicitly describe the expected behaviour, thereby guiding the model toward correct oracles. Similarly, developer-written test cases tend to encode the intended behaviour more effectively than automatically generated test prefixes (e.g., from EvoSuite), which primarily reflect the current implementation rather than the expected one. 

These observations are consistent with prior work~\cite{hossain2025doc2oracll}, which reports that developer-written documentation can improve oracle accuracy by up to 20\%, compared to approximately 10\% improvement when using LLM-generated documentation. Furthermore, this prior work highlights that the effectiveness of documentation depends not merely on its presence, but on its quality: verbose or irrelevant descriptions (e.g., lengthy Javadoc comments without semantic value) do not lead to measurable performance gains.

These observations point to a promising direction for future work: exploring alternative sources of information that more accurately capture expected behaviour. Potential candidates include system-level requirements, commit histories, and bug reports. However, further research is required to systematically assess how such sources can be effectively extracted, represented, and incorporated into LLM-based approaches.

\section{Threats to validity }

\subsection{External Validity}

One potential threat may relate to the subjects we used that we aim at mitigating by selecting datasets that have already been used to evaluate test oracle classification/generation approaches. 
Although our evaluation expands to many tests-assertions pairs and Java projects of different sizes, the results may not generalize to other projects or programming languages. 

Another external threat lies in the tools and LLMs specificity and running configurations we consider. 
To reduce this threat, we employ four of the state-of-the-art commercial and open-source LLMs currently available (GPT 3.5 Turbo, GPT-4o-mini, DeepSeek V3, Codestral), as well as, fundamentally different modern mutation tools and run them using their corresponding default configurations, generating same number of mutants for each subject. 

\subsection{Internal Validity}

Threats to internal validity may arise in the prompts we used. To mitigate this issue, we rely on prompts already used and evaluated by TOGLL \cite{hossain2024togllcorrectstrongtest} and Chattester \cite{yuan2023no}. 
Other threats may relate to how we label the code and assertions as correct or incorrect. To counter this threat, we use a mutation testing tool to inject faults in the correct program version and generate invalid program version, and we use GPT to generate candidate assertions and those that do not pass the test suites were labeled as invalid. In addition, potential threats may be found when decoding LLMs' answers in RQ1 and RQ2. To eliminate this threat, we manually checked that all unique answers given by the models were decoded correctly, and fixed any mistakes found. 

\subsection{Construct Validity}

We mitigate the data-leakage issue by using the Gitbug-Java dataset as part of our experiments, whose code and tests have not been seen during the LLM training, neither the mutants and assertions generation.

Other threat may relate to our assessment metrics. 
However, we employ standard metrics to assess AI-model prediction performances such as accuracy. Instead of reporting general prediction indicators, e.g. precision/recall, we split our evaluation and focus into the four scenarios that allowed us to get specific insights whether the LLM is fed with correct/incorrect code and assertions. 
To assess the test oracle fault detection capabilities, we rely on standard mutation testing techniques and metrics, and we employ one of the state-of-the-art tools to efficiently compute the mutation score.

\section{Related Work}

Several empirical studies have previously evaluated the performance of LLMs in generating complete test suites or test oracles. However, none of these studies have examined whether LLMs mirror the actual implementation or the developer's intent. 

\subsection{Empirical evaluation of LLMs on test generation}

Siddiq et al. \cite{siddiq2023exploring} conducted a study to evaluate LLMs on test generation with focus on compilation rates, correctness, coverage, and test smells. Their evaluation consists of three models: Codex, GPT, and StarCoder. Ouédraogo et al. \cite{ouédraogo2024largescaleindependentcomprehensivestudy} conducted a similar study but by using a broader number of LLMs and incorporated different prompt techniques and strategies to their methodology. While these studies share some common findings regarding the capabilities of LLMs, neither provides a metric to determine whether the generated tests reflect the developer’s intent or simply assert the current implementation. 
Tang et al. \cite{10485640} conducted a study that compares ChatGPT with a non-neural network based method, Evosuite, on terms of correctness, readability, and code coverage. However, they do not compare the two tools on oracle generation or their mutation score capabilities. 
Our study aims at complementing these studies by providing concrete evidence on the LLMs capabilities on test oracle classification and generation, their limitations and challenges for the future research.

\subsection{LLM-based oracle generation and comparisons}

When it comes to the evaluation of LLMs in terms of test oracle generation, there aren't many empirical studies. The first large-scale evaluation of LLMs on test oracle generation comes from TOGLL \cite{hossain2024togllcorrectstrongtest}. Initially the study fine-tunes 7 LLMs and evaluates their performance on test oracle generation. In this study they evaluate the accuracy and the strength of the generated oracles before contributing a tool for generating test oracles. When it comes to the strength of the oracles produced, they evaluate the mutants killed in a complete test suite generated by Evosuite against a complete test suite generated by the pretrained LLMs. In our study, we evaluate the test oracles generated by LLMs against the corresponding Evosuite-generated oracles, focusing on the strength of the test oracles when applied to the same code and, consequently, the same set of mutants. Additionally, our findings suggest that the training data used in TOGLL may limit the potential of LLMs in generating the expected oracles. More importantly, while their study aims to identify an LLM with superior oracle generation capabilities, it does not assess the LLMs' alignment with the developer's intent or compare the classification and generation performance of LLMs.

Apart from the evaluation of LLMs, it worth mentioning that there have been studies that evaluate test oracle generation tools. TOGA evaluates the performance of neural network based techniques, evolutionary algorithms and specification mining methods on the oracles generated \cite{10.1145/3510003.3510141}. However, when future studies evaluated TOGA with different metrics and different datasets \cite{hossain2024togllcorrectstrongtest} have shown that the metrics were unrealistic and that a straightforward baseline was not available \cite{10.1145/3597926.3598080}. Nonetheless, such studies propose and evaluate test oracle generation tools without thoroughly analyzing the underlying LLMs. In this paper, we take a deeper approach, aiming to understand and analyze the behaviour of these LLMs to uncover why these tools perform the way they do. Although both studies exposed limitations of the current LLM-based approaches, none of those studies explains the behaviour of LLMs and how these tools can be improved. 

Since then, more approaches have been used to find the test oracle, each comparing its methodology against others. TECO~\cite{NieETAL22Teco} is a tool that has taken a similar approach to TOGLL. It is based on CodeT5~\cite{wang-etal-2021-codet5} and is finetuned to predict the next statement. Technically, TECO is a tool for test completion, meaning that it is not restricted, nor explicitly trained for oracle generation. However, it has been used and evaluated for its test oracle generation performance on its study, and outperformed the previous state of the art approaches ATLAS and TOGA, as well as the models CodeT5 and CodeGPT.
TECO was evaluated on different metrics than TOGA and TOGLL. Its evaluation was done using lexical-level metrics and functional correctness. For the latter, the compilation success rate and the runnable tests rate were used.

Despite the improvements of the fine-tuned model of TECO against its respective non-finetuned baseline models, its results have been surpassed by ChatAssert~\cite{10804561}, an approach that uses a non-finetuned LLM, GPT-3.5. ChatAssert is a test oracle generation approach that consists of four main features, all of which rely on the capabilities of the LLM: Code summarization, Examples, Dynamic repair and Static repair. Initially, ChatAssert asks the LLM to summarize the code. Then it provides test case examples from the repository under test and asks the LLM to generate an oracle for a given test prefix. Afterwards, ChatAssert compiles the test case and in case of failure it activates its third feature, Dynamic repair. In this phase ChatAssert provides the error message to the LLM and asks it to re-generate the oracle. Once the test compiles, the same procedure is executed on its fourth feature, static repair, which checks whether the test case passes and attempts to regenerate the oracle in case of failure. Additionally, when fixing failed test oracles, ChatAssert keeps a history of failed oracles and provides them as feedback to the LLM in order for the LLM to avoid generating the same responses. 

ChatAssert was not the only approach leveraging the "chain of thought" prompt design for assertion generation. Assertify~\cite{torkamani2024assertifyutilizinglargelanguage} is an approach that uses chain of thought in an attempt to simulate the developer's thinking process. It splits this procedure in four steps. Just like ChatAssert, Assertify's first step is to obtain a summarization from the LLM. In the next steps, it feeds to the prompt more information: the method's input and output and external dependencies. Nevertheless, Assertify is for adding assertions in production code and is not meant for generating test oracles. Hence, there was no evaluation or comparison with test oracle generation techniques. 

Similarly to TECO, AssertT5~\cite{primbs2025assert5testassertiongeneration} is another model based on CodeT5, but specifically trained to generate test oracles. It has been trained on the methods2test dataset~\cite{10.1145/3524842.3528009} and, similar to TECO, outperforms TOGA, ATLAS, and CodeT5. However, it has also been compared to GPT 4o-mini, where it achieves superior results. In addition, AssertT5 has been evaluated on its bug finding capabilities revealing practical limitations on developer-written cases. The Defects4j~\cite{10.1145/2610384.2628055} dataset was used for such an evaluation, where AssertT5 detected only 33 bugs out of the 138 bugs present in the dataset. 

Lastly, it is worth mentioning a test generation approach called CasModaTest~\cite{ni2024casmodatest}. Unlike the previously discussed methods, CasModaTest is a comprehensive test generation framework designed to produce complete test cases, including the test prefix, rather than just test oracles. However, test-oracle generation is an integral part of its internal process. 

The framework consists of two main phases: generation phase and verification phase. The generation phase is split into two components. The first component is responsible for generating the test prefix whereas the second component is responsible for generating the test oracle. During the verification phase, CasModaTest follows a similar approach to the previously mentioned methods, executing the code and leveraging the LLM to resolve compilation issues and fix failed tests. Regarding the test oracle generation process, CasModaTest provides a clear response structure to the LLM, includes relevant code information (method under test, test prefix and human written assertions as a demonstration) and provides a relevant demonstration pool of relevant oracles. The framework utilizes few-shot selection to find the missing oracle. Afterwards, the verification phase takes place which leverages LLMs for fixing the test cases created in the generation phase. 

Lastly, it is worth acknowledging the work of~\cite{molina2025test}, who conducted a literature review of the test oracle problem and the approaches proposed to address it.

\section{Conclusion}

In this study we empirically investigated whether the LLMs can identify the expected program behaviour and thus, be used to classify/generate adequate test oracles. Our findings indicate that LLMs are prone to generate test oracles that capture the actual program implementation rather than the expected one. 
We also observed that developer-like test and variable naming can help the LLM to produce test oracles that capture the expected behaviour, and that the LLM was very effective in discarding invalid assertions.

Results provided evidence that LLMs are more effective in generating valid test assertions rather than in selecting externally generated assertions.
The LLMs managed to produce at least one valid assertion for up to 97.34\% of the test prefixes analysed, which led to higher mutation score than the ones produced by Evosuite. This suggests that incorporating LLMs to generate the test oracles is a promising line of future research.

\bibliographystyle{ACM-Reference-Format}
\bibliography{main}

\end{document}